\documentclass{revtex4}
\usepackage{graphicx}

\usepackage{ulem}
\usepackage[T1]{fontenc}

\begin{document}
\title{Self Similar Properties of Avalanche Statistics in a Simple Turbulent Model}
\author{Roberto Benzi}
\affiliation{Dipartimento di Fisica, Univ. degli Studi di Roma "Tor Vergata", via della Ricerca Scientifica 1, 00133, Roma, Italy}

\author{Ilaria Castaldi}
 \affiliation{Dipartimento di Fisica, Univ. degli Studi di Roma "Tor Vergata", via della Ricerca Scientifica 1, 00133, Roma, Italy}
 
\author{Federico Toschi}
\affiliation{Department of Applied Physics, Eindhoven University of Technology, P.O. Box 513, 5600 MB Eindhoven, The Netherlands}

\author{Jeannot Trampert}
\affiliation{Department of Earth Sciences, Utrecht University, Princetonlaan 8a, 3594 CB Utrecht, The Netherlands}

\keywords{Turbulence, avalanche, intermittency}

\begin{abstract}
In this paper, we consider a simplified model of turbulence for large Reynolds numbers driven by a constant power energy input on large scales. In the statistical stationary regime, the behaviour of the kinetic energy is characterised by two well defined phases: a laminar phase where the kinetic energy grows linearly for a (random) time $t_w$ followed by abrupt avalanche-like energy drops of sizes  $S$ due to strong intermittent fluctuations of energy dissipation. We study the probability distribution $P[t_w]$ and $P[S]$ which both exhibit a quite well defined scaling behaviour. Although $t_w$ and $S$ are not statistically correlated, we suggest and numerically checked that their scaling properties are related based on a simple, but non trivial, scaling argument. We propose that the same approach can be used for other systems showing avalanche-like behaviour such as amorphous solids and seismic events.
\end{abstract}

\maketitle

\section{Introduction}

It is well known that homogenous and isotropic turbulence is characterised by strong intermittent bursts of energy dissipation. An extensive literature exists on the subject and we recommend  \cite{uriel} for an excellent introduction. Turbulence is not the only physical system where energy dissipation occurs intermittently. 

\noindent
Amorphous materials, for example, subject to constant shear rate, show avalanche-like events, which dissipate energy intermittently \cite{fisher}, \cite{barrat}, \cite{schall1}. Analysis of their intermittency mostly focused on the statistical properties of the avalanche size $S$ which usually shows a probability distribution $P[S] \sim S^{-\gamma}$ where $\gamma \in [1.25:1.5] $ depending on the material and its physical properties. Similarly, earthquake dynamics have been discussed in terms of earthquake magnitude or more properly in terms of the seismic moment $M$, which exhibits the celebrated Gutenberg-Richter law $P[M] \sim M^{-{(1+2b/3)}}$, with $b\sim1$ \cite{gr}. Another interesting quantity to consider too is the interevent time $t_w$ between avalanches {\cite{corral2004,benzi1}, which has been poorly investigated in the past. Obviously, both $S$ and $t_w$ depend on the definition of the avalanche, a crucial point we will consider below.  
\bigskip

The concept of avalanche or avalanche-like dynamics is not usually taken into account in turbulence for many reasons: energy dissipation is intermittent both in space and time;  the statistical properties of energy dissipation are usually related to the intermittent fluctuations of energy transfer within the inertial range of turbulence; {an} avalanche does not seem appropriate to discuss energy transfer in the inertial range or in the dissipation range and so on. Thus discussing turbulence in the framework of avalanche dynamics seems at least a rather exotic if not useless approach.  However, we suggest that in some cases the dynamics of turbulent events may be investigated in terms of avalanche dynamics. In this paper, we provide an example based on a very simplified model of turbulence, namely a shell model. For this very special model, we show that avalanche dynamics can be identified and we are able to provide a well-defined meaning to the avalanche size $S$. Not surprisingly, $S$ is related to the rate of energy dissipation. Also, we provide evidence that, for our case, both $t_w$ and $S$ should be considered  {\it bulk quantities}: they are related to the forcing mechanism and do not depend on the statistical properties characterising inertial range fluctuations.  We explore similarities and differences in the statistics of avalanche dynamics in our model compared to results mostly observed in amorphous materials and for earthquakes. Among the similarities, we observe in our shell model a  scaling behaviour of both $S$ and $t_w$ although with different scaling exponents compared to amorphous materials and earthquakes. More importantly, we observe a rather unusual scale invariance in our system which is also observed in other cases. To be precise, we look at the probability distribution $P[t_w|S_{th}]$ of the interevent times $t_w$ for avalanches bigger than some threshold $S_{th}$. Upon increasing $S_{th}$, $P[t_w|S_{th}]$ remains invariant although $t_w$ explicitly depends on $S_{th}$. This is also observed for amorphous materials \cite{benzi1}, \cite{davidsen}, \cite{davidsen2}  and it was first pointed out by Corral \cite{corral2004} by analysing earthquake catalogs; (see \cite{lucilla} for a review).   We also show strong evidence that, if scale invariance holds, then the scaling properties of $S$ and $t_w$ should be related, although both quantities are statistically independent in our model as well in the case of amorphous materials and earthquakes. These relations,  obtained here for the first time, apply {to} our system and amorphous materials despite the difference in the scaling exponents. A speculative conjecture can then be made for earthquake dynamics with excellent agreement with the Corral results.

\bigskip
From the point of view of turbulence or turbulent flows, our approach may be generalised to describe other intermittent behaviours shown by  {\it bulk quantities} whose physical description may be improved by exploiting the same approach outlined in this paper. As an example, we refer to the instability of a thermally driven system in a vertically elongated convection cell.  Such system can be mathematically modelled in terms of a fully periodic thermally driven Rayleigh-Benard cell where the presence of the so-called elevator modes leads to the growth of the kinetic energy associated to vertical motions. This exponential growth is then followed by sudden dissipation events mediated by a shear flow instability that  redistribute such energy horizontally, \cite{federico}. Another example can be found in the study of solar flares which, sometimes, are statistically investigated using models derived in the framework of self-organised criticality \cite{bak}, i.e. in the framework of avalanche dynamics,  see also \cite{vulpiani} for a different point of view.

\bigskip
Our paper is organised as follows: in section \ref{model} we introduce our model and define the forcing mechanism. In section \ref{results} we discuss our definition of event or avalanche size $S$. In the subsections [a] and [b] of section \ref{results} we show that using a different definition of $S$ based on level-crossing, no scale invariance is observed and illustrate the relevance of the forcing mechanism, respectively. All the numerical results are discussed in section \ref{results}.  In section \ref{theory} we provide a theoretical analysis of our system and we show that if scale invariance holds, then there must be a relation between the scaling properties of $t_w$ and $S$. 
We also discuss how our approach can be generalised to amorphous materials and earthquake dynamics with excellent agreement with experimental and/or numerical results. Some general conclusion is provided in section \ref{conclusion}.
We further want to emphasize that our investigation provides a different point of view on turbulent flows complementary to the well-known properties of inertial range dynamics. We suggest that it represents a preliminary step in a new direction which is interesting to study.

\section{Model equations}
\label{model}
We consider a shell model \cite{sabra} to describe our turbulent system  which is defined by the equations:
\begin{equation}
\label{2.1}
\partial_{t} u_{n} = i( k_{n}u_{n+2}u ^{*} _{n+1}-\delta  k_{n-1}u_{n+1}u ^{*} _{n-1} + (1-\delta) k_{n-2}u_{n-1}u_{n-2}) - \nu k_{n}^{2} u_{n}  + f_{n},
\end{equation}
where $u_n$ are complex variables, $k_n = 2^n$ and $\delta = 0.4$. This choice of $\delta$ is know to reproduce the scaling behaviour of $\langle |u_n| ^{p} \rangle \sim k_n^{-\zeta(p)} $ with anomalous exponents $\zeta(p)$ in close agreement with the ones observed in three dimensional turbulence. Thus the statistical properties of the inertial range in the models may be considered close to realistic. The crucial point in our case is the forcing term. We apply $f_n$ to $n=1,2$ (i.e. large scale forcing)  with
\begin {equation}
\label{2.2}
f_n = \frac{A}{u_n^*}  \quad n =1,2
\end{equation}
Using (\ref{2.2}) the rate of energy input $\phi$ in the system is simply given by $4A$ and it is constant. The equation for the kinetic energy $E(t)= \Sigma_n |u_n|^2 $ takes the form:
\begin{equation}
\label{2.3}
\frac{dE}{dt} = \phi - \epsilon(t)
\end{equation}
where $\epsilon(t) = \nu \Sigma_n k_n^2 |u_n|^2$ is the rate of energy dissipation. For the numerical simulations discussed hereafter we chose $\nu=10^{-9}$ and $A = 10^{-3}$. We emphasize  that the following results are independent of the choice of $\nu$ and $A$. By injecting a constant power into our system, the signal for kinetic energy (figure \ref{figura1}, lower panel) shows clearly two different dynamical regimes, one in which the energy grows linearly in time (referred to as "laminar" phase)  and one in which abrupt energy losses occur (referred to as "turbulent" phase). Here laminar and turbulent are used in a  naive way just to distinguish between the two different regimes.
In figure \ref{figura1} (upper panel) we also report the signal corresponding to the energy dissipation for the same time window, which shows, as expected, a strongly intermittent behaviour. Notice the link between the strong fluctuations occurring in the upper signal and the kinetic energy loss in the lower signal.  

\begin{figure}[!h]
\centering
\includegraphics[scale=0.7]{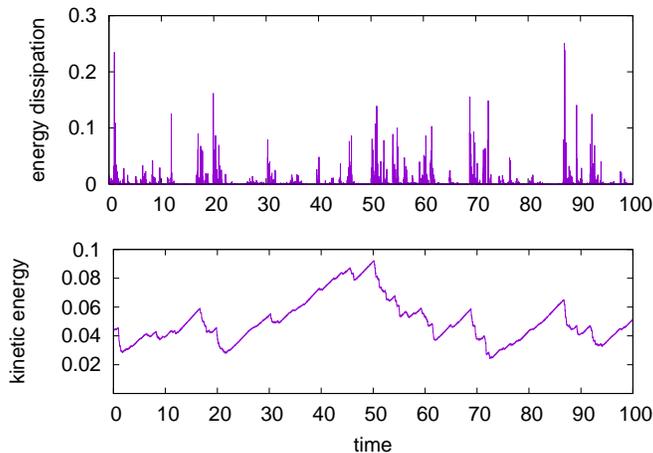}
\caption{Upper panel: rate of energy dissipation $\epsilon(t)$ as a function of $t$ corresponding to eq. (\ref{2.1}) at constant power input. Lower panel: behaviour of the kinetic energy $E(t)$ for the same time window. Looking at $E(t)$,  we can easily distinguish two different dynamical phases in the system: a laminar phase where $E(t) \sim t$ and an avalanche phase where we observe drops of $E(t)$ corresponding to the bursts in the energy dissipation. }
\label{figura1}
\end{figure}

\section{Event Definition and Scale Invariance}
\label{results}
Having in mind the signal shown in figure (\ref{figura1}),  we define an avalanche as the event for which $dE/dt < 0$. The starting point of an event is identified by a change in the sign of the derivative of kinetic energy in time, from positive to negative, meanwhile, the opposite change in sign of $dE/dt$ identifies the end of the avalanche or energy drop. In figure (\ref{figura2}) we illustrate our definition.
For each event we compute the avalanche size as $ S(k) = - \int _{t_{i}(k)} ^{t_{f}(k)} dt (dE/dt)$, where $k$ is a label for the event ($k=1,2...$),  $t_i(k)$ is the initial time of the $k$ event  and $t_f(k)$  its corresponding final time. For $t \in [t_i(k),t_f(k)]$ we require that $dE/dt<0$. Thus $t_i(k)$ is identified by the condition $\epsilon(t_i(k)) > \phi$ and $t_f(k)$  by the condition $\epsilon(t_f(k))\le \phi$. $S(k)$ thus represents the total energy release during avalanche event $k$.

\bigskip

Having thus  defined avalanches, we can look at the probability distributions of the avalanche size $S$ and the interevent times $t_w$ The latter is defined as $t_w(k) = t_i(k)-t_f(k-1)$. In figure (\ref{figura3}) we shows $P[S]$ (right panel) and $P[t_w]$ (left panel): both probability distributions show  very clear scaling regions  over several decades. We indicate the corresponding scaling exponents  by $\gamma$ and $\alpha$:
\begin{equation}
    P[t_{w}] \sim \frac{1}{t_{w}^{\alpha}} \quad P[S] \sim \frac{1}{S^{\gamma}}
\end{equation}
A best fit estimate gives $\alpha=1.65$ and $\gamma = 0.35$ with an accuracy of the order of a few percent.

\bigskip

\begin{figure}[!h]
\centering
\includegraphics[scale=0.6]{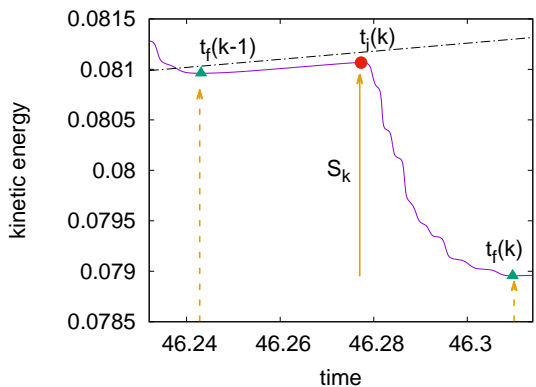}
\caption{The figure illustrates our definition of event. We define an avalanche $S(k)$, $k=1,2,..$,  as the event in time for which $dE/dt<0$ in the time interval 
$t \in [t_i(k),t_f(k)]$ where  the times $t_i(k)$ and $t_f(k)$ corresponding to the initial time and final time of the avalanche. The interevent time $t_w(k)$ is indicated by the two arrows: $t_w(k) = t_i(k)-t_f(k-1)$. }
\label{figura2}
\end{figure}

\begin{figure}[!h]
\centering
\includegraphics[scale=0.5]{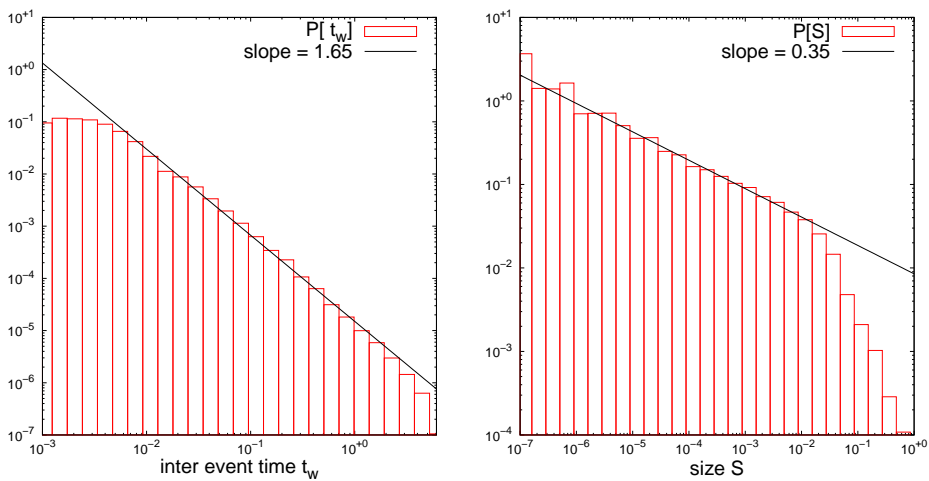}
\caption{Probability distribution of the interevent time $P[t_w]$ (left panel) and avalanche size $P[S]$ (right panel) observed for modelling eq. (\ref{2.1}) with a constant power input. In both cases well defined scaling laws are observed with $P[t_w] \sim t_w^{-1.65}$ and $P[S] \sim S^{-0.35}$. The accuracy in the exponents are of the  order of  $5\%$.}
\label{figura3}
\end{figure}
It is important to understand that both $t_w$ and $S$ are not trivially linked to the scaling properties of the inertial range fluctuations in the system. The value of $t_w$ is dictated by the onset of some instability occurring at relatively large scales whereas the avalanche of size $S$ depends on the short intermittent bursts of energy dissipations. Although the initial and the final times of an avalanche occur when $\epsilon(t) >\phi$ and $\epsilon(t)<\phi$, respectively, its size does not necessarily correspond to large or small values of $\epsilon-\phi$. In other words, large values of $\epsilon$ can occur for both small and large values of the avalanche size $S$. 
More precisely the size $S(k)$ of the avalanche can be computed  from eq.(\ref{2.3}) to give $S(k) = \int_{t_i(k)}^{t_f(k)} \epsilon(t) dt - \phi (t_f(k)-t_i(k))$ with $\epsilon(t) \ge \phi$ for $ t \in [t_i(k),t_f(k)]$. Upon denoting the event duration by $\tau(k) = t_f(k)-t_i(k) $,  the probability distribution of  $S(k)$   can be obtained from the probability distribution   of $ \tilde \epsilon(\tau(k)) \equiv  \int_{t_i(k)}^{t_f(k)} \epsilon(t) dt $ constrained by the previously mentioned conditions on $\epsilon(t)$. Furthermore $\tilde \epsilon (\tau(k))$ should be computed in the dissipation range and its fluctuations are correlated (with some non trivial lag time depending on $\tau(k)$) to the probability distribution of the large scale velocity fluctuations, i.e. to the kinetic energy. In summary,  besides the fact that the statistical properties of $S(k)$ depend on the inertial range fluctuations in a complicated way, the probability distribution of $P[S]$ should be self-consistent with the fluctuations of the kinetic energy. Thus, the knowledge on the statistical properties of the inertial range velocity fluctuations does not provide any short cut  to estimate the probability distributions of both $t_w$ and $S$.

\bigskip

We shall see later (in subsection [b]) that the statistical properties of both $t_w$ and $S$ depend on the way we force our system, i.e. on the physical mechanism of energy input in the system. From this point of view, both $t_w$ and $S$ may be considered as {\it bulk quantities} which characterise the random dynamics of the energy behaviour with respect to the (given) external forcing. There is no reasons {\it a priori} for $P[t_w]$ and/or $P[S]$ to be scaling functions of their arguments. The results shown in figure (\ref{figura3}) are therefore non trivial.

\begin{figure}[!h]
\centering
\includegraphics[scale=0.8]{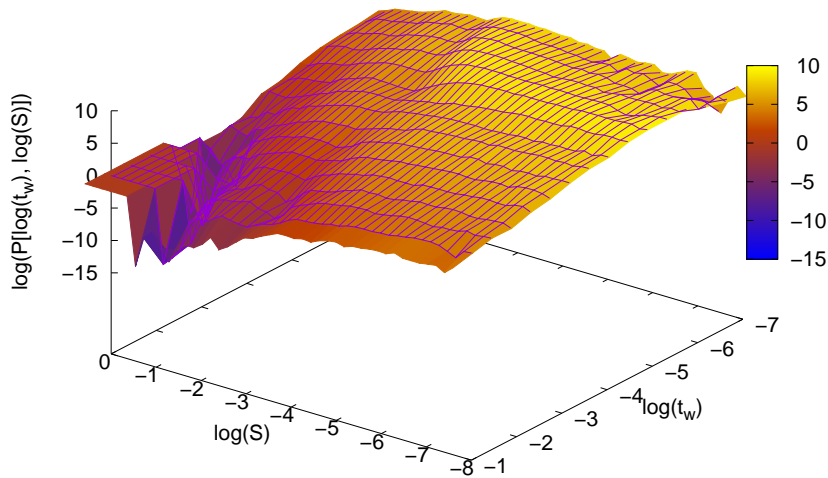}
\caption{In the figure,  we show a 2D map for the PDF of both $log(S)$ and $log(t_{w})$.  
The figure highlights the statistical independence of the avalanche size and the interevent times.  
By looking at the figure, there is no evidence of any correlation between the two variables, and by a numerical check, their correlation was found of the order of 7\%.}
\label{figura4}
\end{figure}

\bigskip
An important observation is that there exists no statistical correlation between $t_w$ and $S$. In figure (\ref{figura4}) we show the joint probability distribution of $t_w$ and $S$  which does not provide any significant hint of correlations between $t_w$ and $S$. The statistical independence of $t_w$ and $S$ is also observed in other systems, such as amorphous solids and/or soft glasses, where avalanche dynamics are also characterised by scaling functions of $P[t_w]$ and $P[S]$, albeit with completely different scaling exponents \cite{benzi1} \cite{davidsen2}.  Furthermore,  a detailed analysis of seismic events \cite{corral2006} shows that no significant correlations exist between their interevent times and earthquake magnitudes. Thus, the statistical independence of $t_w$ and $S$ is a rather common feature observed in systems characterised by avalanche-like dynamics. 

\bigskip

Following \cite{corral2004}, we investigate a rather intriguing property of the interevent time statistics. Using our definition of avalanche, its size $S$ spans from some minimum, say $S_0$, to some maximum $S_M$. Analogously the interevent times $t_w$ spans from some minimum $t_0$ and some maximum $t_M$. Let us now consider the interevent times $t_w$ occurring for avalanches greater than $\lambda S_0$, where $\lambda$ is some real positive number. The interevent times are modified as illustrated in figure (\ref{figura5}).  The figure shows the energy behaviour $E(t)$ during a relatively short time window: the continuous line represents $E(t)$ and the red points highlight the avalanches occurring during the selected time window. One can see two relatively large avalanches occurring at the begging and at the end of the time window with rather small events in between. Here large and small refer to the avalanche size. We also highlight the interevent times between avalanches. Once we consider avalanches bigger than $\lambda S_0$ ($\lambda>1$), for some value of $\lambda$,  the two avalanches in the middle are neglected and the interevent time (corresponding now to the dashed line in the figure) becomes longer and is approximatively equal, in this particular example, to $t_1+t_2+t_3$. Obviously, the probability distribution $P[S]$ remains unchanged and we may further assume that $t_0$  remains unchanged as well as $S_M$. However, clearly,  this is not the case for $t_M$,  which we expect to increase. Thus we should wonder how the probability distribution of the interevent times changes. We denote this new probability distribution by $P[t_w | S_{th}]$ where $S_{th}= \lambda S_0 $ is the size of the avalanches disregarded for the computation of $t_w$.

\bigskip
In figure (\ref{figura6}) we show $P[t_w | S_{th}]$ for different values of $S_{th}$ spanning almost two orders of magnitude.  A rather striking results is observed: $P[t_w|S_{th}]$ is invariant,  i.e. it is a scaling function with the same exponent $\alpha \sim 1.65$. We notice that the very same results are observed in the interevent times of avalanches for amorphous materials \cite{benzi1} \cite{davidsen2} and earthquakes  \cite{corral2004} \cite{lucilla},  although  $\alpha<1$ and the scaling range is smaller compared to what is observed in figure (\ref{figura6}). In the model presented in this paper,  the invariance of $P[t_w | S_{th}]$  is  quite clear and striking. $P[t_w|S_{th}]$ can therefore  be considered invariant with respect to the transformation $S_0 \rightarrow \lambda S_0$. 

\bigskip

\begin{figure}[!h]
\centering
\includegraphics[scale=0.6]{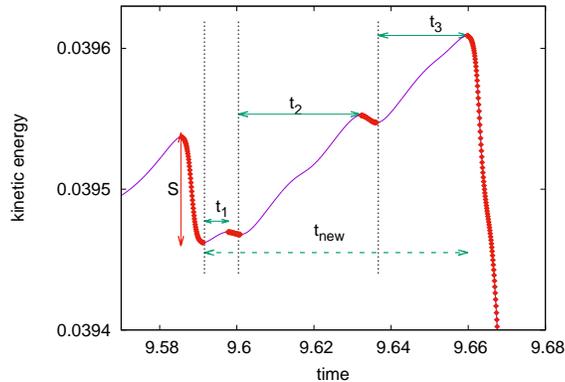}
\caption{The figure illustrates the change in the quantitative definition of the interevent time by looking at avalanches whose sizes $S$ are larger than some threshold $S_{th}$. The continuous line corresponds to $E(t)$. The red dots indicates when $dE/dt<0$. We observe $4$ avalanches. The vertical continuous arrow shows the size $S$ of the first avalanche on the left. If we assume the two middle avalanches to be smaller than $S_{th}$, the interevent time between the first and the next avalanche becomes $t_1+t_2+t_3$.  }
\label{figura5}
\end{figure}

What is the physical meaning behind the results shown in figure (\ref{figura6})? This is a non-trivial question which we will try to answer qualitatively. The scaling transformation $S_0 \rightarrow \lambda S_0$ should be considered equivalent to a kind of {\it coarse grained}  transformation in the system: upon considering larger avalanche sizes we study the dynamics of the system on a {\it longer}  time scale. Thus we may consider the invariance of $P[t_w| S_{th}]$ as the signature of a {\it scale invariance} of the system dynamics. This would be  somehow trivial {\it if} $P[t_w|S_{th}]$ were a scaling function of both $t_w$ {\it and} $S_{th}$. However, this is not the case in our system (nor in amorphous materials nor for earthquakes) simply because interevent times are statistically independent of the avalanche sizes. Thus we are looking at a rather peculiar case of scale invariance,  which deserves a deeper investigation. In section [\ref{theory}] we make the first step in this direction.

\begin{figure}[!h]
\centering
\includegraphics[scale=0.6]{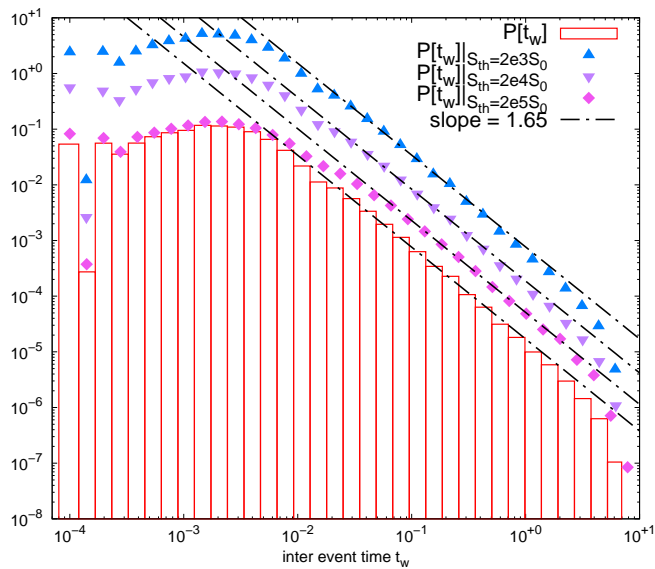}
\caption{The figure illustrates the behaviour of the probability distribution $P[t_w|S_{th}]$ which is the probability distribution of the interevent times $t_w$ occurring between avalanches of size $S>S_{th}$.   Here we show the results for three different value of $S_{th}$ which is defined with respect to the smallest size we found in our system,$S_{0}$, by $S_{th}=\lambda S_{0}$. Upon changing $S_{th}$ for over two orders of magnitude the probability distribution of the interevent times preserves its scaling behaviour with the same scaling exponents as shown in figure (\ref{figura3}). }
\label{figura6}
\end{figure}

\subsection{A Different Statistical Analysis}

First, we illustrate how the definition of $P[t_w | S_{th}]$  gives different results using different approaches usually employed in the analysis of intermittent or random processes. In particular, we consider a rather common approach based on the statistical properties of level-crossing. Given the dynamics of energy dissipation $\epsilon(t)$ one can define an avalanche event as the time interval $[t_i(k),t_f(k)]$ for which $\epsilon(t) > \epsilon_*$. The size $S$ of the event is defined as $S(k) = \int_{t_i(k)}^{t_f(k)} \epsilon dt$ while the interevent times are defined as usual $t_w(k) = t_i(k)-t_f(k-1)$. This definition of event refers to  {\it level-crossing} because it depends on $\epsilon_*$.  The statistical properties of level crossing are relevant quantities worthwhile to investigate in many physical and mathematical problems and there exists an extensive literature on the subject \cite{adler} \cite{santucci}.  In the framework of level-crossing, both $P[S]$ and, more importantly, $P[t_w]$ depend on  $\epsilon_*$. We now use this approach using our model and the very same data set as employed in figure (\ref{figura6}). In particular, we focus on the probability distribution of the interevent times $P[t_w| \epsilon_*]$ for different values of $\epsilon_*$ in figure (\ref{figura7}). Different to figure (\ref{figura6}) $P[t_w|\epsilon_*]$ is no longer invariant, i.e. it is still  a scaling function of $t_w$ but with  scaling exponents $\alpha$ which decrease upon increasing $\epsilon_*$. We argue that the different results shown in figures (\ref{figura6}) and (\ref{figura7}) can be explained by noticing that upon increasing $\epsilon_*$ we are changing the physical meaning of the event. 
This is because in the $\epsilon$ based definition of size, the term $\phi (t_f(k)-t_i(k))$ is missing.
Consider,  for instance, a relatively large event selected with the level-crossing at some value  $\epsilon_*$. Upon increasing $\epsilon_*$ the initial time of the event is shifted and the corresponding interevent time increases or event duration decreases. This  implies that $t_w $ and $S$ may acquire non-negligible correlations for large enough $\epsilon_*$.  On the contrary, the scale invariance shown in figure (\ref{figura6}) is based on the same definition of the event regardless of the threshold $S_{th}$. It is true that the definition of event used in figure (\ref{figura6})  is also based on a threshold, namely $\phi$, however, looking at the interevent times for avalanches $S>S_{th}$ we retain the same definition (i.e. the events for which $\epsilon(t) \ge \phi$) while neglecting in the computation of $t_w$ avalanches smaller than $S_{th}$.  Thus $P[t_w|S_{th}]$ and $P[t_w|\epsilon_*]$ refers to two different statistical properties of the system and are not related to one another. 
This example highlights the fact that the scale transformation $\epsilon_* \rightarrow \lambda \epsilon_*$ cannot be considered a {\it coarse-grained } transformation in the sense previously discussed. We are still looking at longer time scales but we are also considering different events. This is a crucial point often not properly taken into account in various statistical analysis.

\bigskip

Finally, let us remark that the results shown in figure (\ref{figura7}) beautifully illustrate that the property of scale invariance shown in figure (\ref{figura6}) is not trivially linked to the probability distribution of $\epsilon(t)$ which is the same for both cases.

\begin{figure}[!h]
\centering
\includegraphics[scale=0.6]{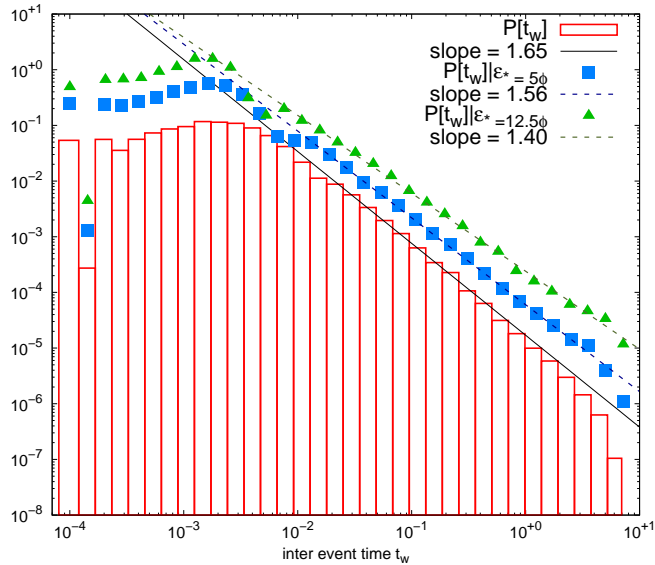}
\caption{In this figure, we compute the probability distribution of $P[t_w, \epsilon_*]$ which represents the probability distribution of interevent times between consecutive events where $\epsilon(t) > \epsilon_*$ where $\epsilon(t)$ is the rate of energy dissipation and $\epsilon_*$ is defined with respect to the constant injected power $\phi = 4A$. Contrary to what we observe in figure (\ref{figura6}), the probability distribution $P[t_w,\epsilon_*]$ is not invariant upon increasing $\epsilon_*$. }
\label{figura7}
\end{figure}

\subsection{A Different Forcing}

\begin{figure}[!h]
\centering
\includegraphics[scale=0.6]{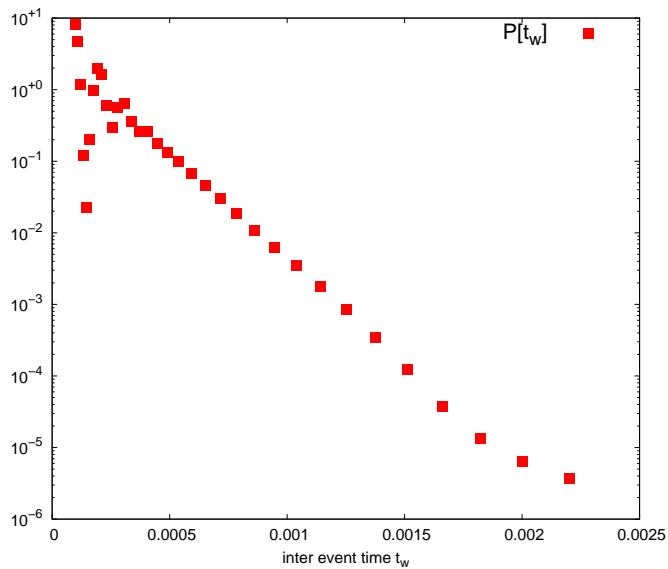}
\caption{Probability distribution $P[t_w]$ of the interevent times $t_w$ obtained for a random forcing acting on the shell $1$. Notice that $P[t_w]$ does not show any scaling behaviour and it can be approximated as an exponential distribution.
}
\label{figura8}
\end{figure}
Another important issue to clarify is the relevance of the large-scale forcing. We have previously argued that the statistical properties of both $t_w$ and $S$ should be considered as  {\it bulk} quantities of our system, i.e. quantities linked to the way the system is forced and describing the dynamics {\it with respect to the forcing mechanism}. We now considered exactly the same model given by eq. (\ref{2.1}) but  we assume $f_n$ to be a random gaussian process $\delta$ correlated in time and acting for $n=1$.   We use the same definition of event size introduced in the first part of this section, namely $S(k) = - \int _{t_i(k)}^{t_f(k)} dt(dE/dt)$  and we look at the probability distribution of $P[t_w]$ and $P[S]$. 

There is no evidence of $P[t_w]$ to be a scaling function of $t_w$, see figure (\ref{figura8}). Actually, we observe that $P[t_w]$ can be very well approximated by an exponential distribution. 
Notice that the scaling properties of the shell model, i.e. the anomalous scaling $\langle |u_n||^p \rangle \sim k_n^{-\zeta(p)}$, remains unchanged using the stochastic forcing. Thus, figure (\ref{figura8}) illustrates the point raised at the beginning of this section, namely that the statistical properties of $t_w$ and $S$ are not linked to the inertial range dynamic, but a different forcing gives different statistical properties of the avalanche events. This also implies that scale invariance shown in figure (\ref{figura6}) is not due to the scaling properties of the inertial range dynamics.

\section{A tentative theory for scaling exponents}
\label{theory}

We will  provide a theoretical framework to discuss the results illustrated in section [\ref{results}]  and in particular the property of scale invariance of $P[t_w|S_{th}]$ shown in figure (\ref{figura6}).  To fix our theoretical analysis, we assume that  the probability distributions of the  avalanche size $S$ and interevent time $t_w$ are scaling functions of their arguments, i.e.
\begin{eqnarray}
\label{t1}
P[S] = \frac{Z_S}{S^{\gamma}} \,\,\,\, S \in [S_0, S_M]  \\
\label{t2}
P[t_w] = \frac{Z_t}{t^{\alpha}} \,\,\,\, t_w \in [t_0,t_M] 
\end{eqnarray}
where $Z_S$ and $Z_t$ are normalization factor. Notice that we focus only on the scaling part of $P[t_w]$ and $P[S]$ neglecting  regions where no scaling is observed. For our analysis this approximation is  reasonable.  Upon assuming $t_M \gg t_0$ and $S_M \gg S_0$, for $\gamma <1 $ and $\alpha >1$ (as in our case) we have to the leading order $Z_S = S_M^{\gamma-1} $ and $Z_t = t_0^{\alpha-1}$. 

\bigskip
We now consider the quantity:
\begin{equation}
\label{t3}
 X \equiv \frac{ E_{stored}}{E_{relesased}}
 \end{equation}
where $E_{stored}$ is energy input in the system during the interevent time and $E_{released} = S$ is the energy released during an avalanche. Because we are forcing the system with a constant power input $\phi = 4A$, $E_{stored} $ is well approximated  by the quantity $4A t_w$. Hereafter, we can disregard the factor $4A$ and we  write
\begin{equation}
\label{t4}
X = \frac{t_w}{S}
\end{equation}
 $X$ is a random variable that describes the fluctuations of the dynamical process in the energy behaviour of the system, namely the stored energy in the system with respect to the energy released.  Now, let us consider the scale transformation
\begin{equation}
\label{t5}
S_0 \rightarrow \lambda S_0
\end{equation}
As we argued in  section [\ref{results}] , the scale transformation (\ref{t5}), does not change $t_0$ whereas $S_M$ is not changed by definition. Because we expect $t_M$ to increase, we can write

\begin{equation}
t_M \rightarrow \lambda^H t_M
\label{t6}
\end{equation}
where $H$ is yet unknown. In general we expect that $P[t_w]$  will change because of the scale transformation (\ref{t5}) and the exponent $\alpha$ may become a function of $\lambda$, i.e.
\begin{equation}
\label{t7}
P[t_w] =  \frac{t_0^{\alpha(\lambda)-1}}{ t_w^{\alpha(\lambda)}}
\end{equation}

\bigskip
We are interested in the probability distribution $P[X]$. Under the scale transformation (\ref{t5}) we expect that  $X$ and $P[X]$ depend on $\lambda$. The scale transformation (\ref{t5}) can be interpreted as a "coarse grained" transformation following our discussion in the  section [\ref{results}]. From this point of view, scale invariance in the system should be equivalent to saying that $P[X]$ does not depend on $\lambda$. Because, $t_w$ and $S$ are independent variables, we can easily compute the moments $\langle X^{-n} \rangle$  and $\langle X^n \rangle$ for any $n$, where $\langle .. \rangle $ is the average over $P[X]$. Using (\ref{t1}) and (\ref{t7}) we obtain:
\begin{equation}
\label{t8}
\langle X^{-n} \rangle =  \left[ \frac{S_M}{t_0} \right]^{n}
\end{equation}
Since neither $t_0$ nor $S_M$ depends of $\lambda$,  we obtain that $\langle X^{-n} \rangle$ is independent on $\lambda$. To simplify the following computation, we can assume $t_0=1$ and $S_M=1$ without loss of generality.  To find  $\langle X^n \rangle$ with $n>0$, a little algebra  gives:
\begin{equation}
\label{t8}
\langle X^n \rangle = t_M^{n - \alpha(\lambda)+1} S_0^{1-n-\gamma}
\end{equation}
Under the scale transformation (\ref{t5}) we then obtain
\begin{equation}
\label{t9}
\langle X^n \rangle \rightarrow  \langle X^n \rangle \lambda^{ [n(H-1) +1+H -H\alpha(\lambda)-\gamma]}
\end{equation}
For scale invariance to hold we must require that for any $n>0$ the following equation is satisfied:
$$
n(H-1) +1+H -H\alpha(\lambda)-\gamma = 0
$$
This implies that $H=1$, $\alpha(\lambda)$ is independent of $\lambda$ and:
\begin{equation}
\alpha + \gamma = 2
\label{t10}
\end{equation}
This tells us something interesting: first of all, the scale invariance of $P[X]$ is equivalent to the scale invariance of $P[t_w|S_{th}] ]$ under the scale transformation (\ref{t5})  and $\alpha$ does not depend on $\lambda$; secondly, we obtain a non-trivial result relating the scaling exponents $\gamma$ and $\alpha$ and expressed by eq. (\ref{t10}). It is important to note that our results do not "prove" scale invariance in our system. What we can prove is that {\it if the statistical properties of the system are scale-invariant with respect to (\ref{t5}) then the scaling exponents $\alpha$ and $\gamma$ are not independent and satisfy eq. (\ref{t10})}.
The numerical values of $\alpha $ and $\gamma$ obtained in the previous section, see figure (\ref{figura3}),  are in excellent agreement with (\ref{t10}) within few percent. As a side product of eq. (\ref{t10}) we observe that the probability distribution of $\Xi= 1/S$ should be a scaling quantity with scaling exponent $\alpha$. Using this observation, we can provide a direct test of eq. (\ref{t10})  by comparing the probability distribution of $1/S$ with respect to the probability distribution $P[t_w]$. 
This is done in figure (\ref{figura9}) which shows an excellent agreement with (\ref{t10}). Thus, besides the numerical estimate of $\alpha$ and $\gamma$, we can check the validity of eq. (\ref{t10}) directly in figure(\ref{figura9}).
\begin{figure}[!h]
\centering
\includegraphics[scale=0.6]{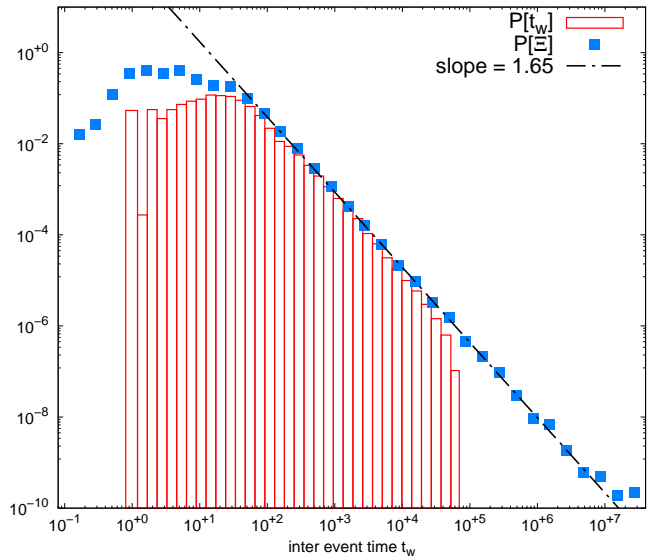}
\caption{Quantitative check of our theoretical approach.  According to our theoretical interpretation of scale invariance, the probability distribution of $\Xi \equiv 1/S$  should show the scaling properties of the interevent time probability distribution $P[t_w]$. This is checked in the figure where we compute $P[\Xi]$.}
\label{figura9}
\end{figure}

\bigskip

It is  tempting to investigate whether the same argument can be applied {to} other systems. In a soft glass and/or in an amorphous solid driven by a constant shear rate $\dot \Gamma$, the internal energy grows as $\sigma \Gamma$ where $\sigma$ is the internal stress and $\Gamma = \int \dot \Gamma dt $ is the applied strain. Since $\sigma \sim \Gamma$ the internal energy of the system, during the time between two consecutive avalanches, grows as $t_w^2$ under the (usual) assumption that $\dot \Gamma$ is constant. For the systems  where both $\gamma$ and $\alpha$ have been measured \cite{benzi1} \cite{davidsen2}  ($ \alpha<1$ and $\gamma \sim 1.33 >1$),  the same reasoning introduced above in this section gives:
\begin{eqnarray}
\label{t11}
\langle X^{-n} \rangle &=& \left [\frac{t_M^2}{S_0} \right]^n \\
\label{t12} 
\langle X^n \rangle &=& \left[ \frac{S_0}{S_M} \right]^{\gamma-1} S_M^n \left[ \frac{t_M}{t_0} \right]^{\alpha} \frac{1}{t_M t_0^{2n-1}}
\end{eqnarray}
Eq. (\ref{t11}) tells us that upon scaling $t_M \rightarrow \lambda^{1/2} t_M$ (i.e.$H=1/2$ ) for scale invariance to hold. Eq. (\ref{t12}) then implies that:
\begin{equation}
\label{t13}
\alpha = 3-2\gamma
\end{equation}
For $\gamma \sim 1.33$ we obtain $\alpha  \sim 0.33 $ in very good agreement with the numerical and experimental results discussed in \cite{benzi1}. Thus, it seems that our argument can be considered rather general and independent of the detailed physical mechanisms behind the avalanche dynamics. 

\bigskip

The common assumption in deriving eq. (\ref{t10}) and (\ref{t13}) is that the system is scale-invariant under the transformation (\ref{t5}). Physically,  in both cases,  we can argue that the system approaches some kind of  {\it critical dynamics} where two different phases (laminar/turbulent in our case) or (no flow/flow for amorphous systems) are dynamically competing. The scale invariance,  {\it if it occurs }, then implies that the scaling exponents of avalanche size and interevent times are linked. This is a rather non-trivial outcome of our analysis.  

\bigskip

{It is equally tempting} to address the case of earthquakes where the Gutenberg-Richter scaling implies $\gamma = 5/3$. For earthquakes, however, we have no idea how stored energy depends on the time between two consecutive events.  For $\gamma=5/3$ we have $\alpha = 0.33$ {using(\ref{t10})} in agreement with the results obtained in \cite{corral2004} and, interestingly, close to the one obtained in amorphous solids. { This suggests that the stored energy grows linearly in time. How reasonable is this? We can speculate that energy stored is still given by $\sigma \Gamma$ as in amorphous materials. Then, following \cite{Madariaga} we can assume that  $\sigma$ is equal to the so-called {\it apparent stress},  which is commonly assumed to be constant. The strain  $\Gamma$ is due to tectonic motion and, in a very first approximation, we may argue that it is proportional to time. Using these (strong) assumptions on the stress and the strain,  we can write $X \sim t_w /S$ and we can repeat the same reasoning leading to (\ref{t10}), which leads of course to the same relation $\alpha + \gamma = 2$.}. Although exciting, we should consider our finding very preliminary and very speculative and the argument definitively deserves more rigorous investigations.

\bigskip

\section{Conclusion}
\label{conclusion}

This paper  discusses avalanche dynamics in a shell model of turbulence forced with a constant power input $\phi$. The  avalanche-like events are characterised by {a} sharp negative decrease of the kinetic energy $E$. We investigated two relevant statistical properties, namely the probability distribution $P[S]$ of the avalanche size $S$, corresponding to the energy drop during an event, and the probability distribution $P[t_w]$ of the interevent time $t_w$  between $2$ consecutive avalanches. Both probability distributions show a  clear scaling behaviour $P[S] \sim S^{-\gamma}$,  $\gamma = 0.35$,  and $P[t_w] \sim t_w^{-\alpha}$,  $\alpha = 1.65$. We have provide numerical evidence  that the probability distribution of $t_w$ shows  {\it scale invariance}: upon computing $t_w$ between events of size $S \ge S_{th}$, $P[t_w]$ shows the same scaling behaviour independently of $S_{th}$, while  Importantly $t_w$ and $S$ are statistically independent variables. 

\bigskip
This scale invariance is similar to what has been observed in amorphous materials and in the analysis of earthquake catalogs. Assuming scale invariance to hold, we provide a simple theoretical argument stating that the scaling exponents $\alpha$ and $\gamma$ must satisfy the relation $\alpha+\gamma=2$ in excellent agreement with the numerical results. We have generalised our approach for amorphous materials with very good agreement against numerical and experimental results, 
giving $\alpha + 2\gamma = 3$. Relations (\ref{t10}) and (\ref{t13}) are here derived for the first time. Once again, we remind {the reader} that, in our view, both relations are consequences of scale invariance in the system, and they don't necessarily hold for all systems showing avalanche-like dynamics. For instance, in \cite{benzi1} it was shown that the statistical properties of $t_w$ in amorphous materials depend critically on the material stiffness. In particular, scale invariance is observed for very "rigid" systems, whereas this is not true for softer materials. When scale invariance holds, then the scaling exponents of $P[t_w]$ and $P[S]$ are related {to one another}. This is something new and somehow unexpected. 

\bigskip
One important point highlighted in the paper is that neither $P[S]$ nor $P[t_w]$ can be obtained using statistical properties of inertial range (intermittent) fluctuations. For this reason, we consider both $P[S]$ and $P[t_w]$ as {\it bulk quantities} related to the forcing mechanism. An important possible question to address is whether the scaling exponents depend on the $Re$ number.
\bigskip

\begin{figure}[!h]
\centering
\includegraphics[scale=0.6]{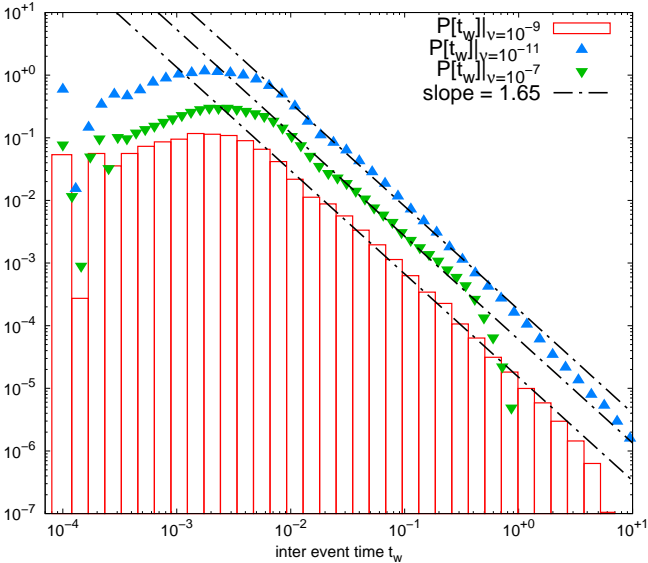}
\caption{Probability distribution $P[t_w]$ of the interevent times $t_w$ obtained eq. (\ref{2.1}-\ref{2.2}) for different values of $\nu$. Upon changing the $Re$ number, the scaling properties of the $P[t_w]$ does not change.  }
\label{figura9}
\end{figure}

We numerically checked that neither $\alpha$ nor $\gamma$ are function of $Re$. In figure (\ref{figura9}) we show $P[t_w]$ for $\nu = 10^{-7}$,$10^{-9}$ and $10^{-11}$: upon changing $Re$ by $4$ orders of magnitude, the scaling exponent $\alpha$ does not change (the same is true for $\gamma$, not shown). The only relevant change in $P[t_w]$ concerns the range where scaling is observed which seems to decrease with $Re$.   

\bigskip

Another important point is how to properly define scale invariance for the probability distribution $P[t_w]$. Interevent times (sometimes referred to as return times or waiting times) are important statistical variables discussed for many theoretical frameworks in physics. Given a random (intermittent) process $v(t)$, one can study $P[t_w,L] $ as a function of the time at which $v(t)$ is larger than some level $L$. In the case discussed in this paper, $P[t_w, L]$ does not show any scale invariance. In fact, upon changing $L$, we are also changing our definition of {\it event size}. Our approach, following \cite{corral2004}, is to fix the definition of event and then disregard it in the computation of $t_w$ events of size larger than some threshold $S_{th}$: we are looking at some longer time for the {\it same} set of events. One may argue that this is a minor detail {, but} we have clearly shown that this is not the case in our system and we argue the same is true for other physical systems. 

\bigskip
There is no {\it a priori} reason to assume that the interevent time distribution $P[t_w]$ is scale-invariant in the sense discussed in this paper (see also the discussion at the end of section [\ref{theory}] and \cite{benzi1}).  However, it is a remarkable result, here presented for the first time, that if scale invariance holds then the scaling exponents of size and interevent time distributions must be related.

\vskip6pt

\enlargethispage{20pt}

Data are accessible at the address: https://doi.org/10.4121/14546958. All authors made equal contribution to the paper. All authors read and approved the manuscript. The author(s) declare that they have no competing interests.\\

Acknowledgments: This paper is dedicated by two of us (RB and FT) to our colleague and friend Prof. Uriel Frisch. Along the years, Uriel has been an invaluable font of inspirations in our research work and we are grateful for  his tireless encouragement, advice and his extraordinary sense of humor.\\


\end{document}